# Correlation of Magnetic State Configurations in Nanotubes with FMR spectrum

Abhishek Kumar[1]#, Chirag Kalouni[1]#, Raghvendra Posti[1], Vivek K Malik[2], Dhananjay Tiwari[3], and Debangsu Roy[1]*

[1]*Department of Physics, Indian Institute of Technology Ropar, Rupnagar 140001, India*

[2]*Department of Physics, Indian Institute of Technology Roorkee, Roorkee 247667, India*

[3]*Silicon Austria Labs GmbH, Sandgasse 34, Graz 8010, Austria*

## Abstract

Magnetic nanotubes have garnered immense attention for their potential in high-density magnetic memory, owing to their stable flux closure configuration and fast, reproducible reversal processes. However, characterizing their magnetic configuration through straightforward methodologies remains a challenge in both scope and detail. Here, we elucidate the magnetic state details using Remanence Field Ferromagnetic Resonance Spectroscopy (RFMR) for arrays of electrodeposited nanotubes. Micromagnetic simulations revealed distinct spin configurations while coming from saturation, including the edge vortex, onion, uniform and curling states, with chirality variations depending on the preparation field direction. Dynamic measurements, coupled with RFMR spectra analysis, unveiled multiple FMR modes corresponding to these spin configurations. The evolution of spin configurations under bias fields were studied, indicating nucleation within the curling state. Observations revealed opposite RFMR spectra, denoting opposite magnetic spin configurations after removing the positive and negative saturating fields when the magnetic field was applied along ($\theta_H = 0^o$) and perpendicular ($\theta_H = 90^o$) to the nanotube axis. We observed a mixture of the non-uniform curling states with the end vortex state (onion-like curling state) at the end of the nanotubes for the $\theta_H = 0^o$ ($90^o$) and uniform magnetization states in the middle of the nanotubes for the $\theta_H = 0^o$ configuration. Building on RFMR information, frequency-swept FMR absorption spectra obtained at different bias fields allowed the characterization of magnetization states. This picture was supported by micromagnetic simulations. These findings were further substantiated with First Order Reversal Curve measurements (FORC).

# Both the authors contributed equally.

* Corresponding author: debangsu@iitrpr.ac.in

# Introduction

The exploration of magnetic nano conduits is driven by their potential as future high-density magnetic memory[1, 2]. It is envisioned that the information in such devices will be encoded through the propagation of domain walls, induced by spin torque and/or spin-polarized current in the magnetic track[3, 4]. Additionally, the stray field-free states ensure higher magnetic element density, while the speed at which domain walls can be propagated in the conduits dictates the device's speed[5]. In this regard, ferromagnetic nanotube becomes important due to its lack of magnetic core resulting in stable flux closure configuration and fast and reproducible reversal processes[6, 7]. At equilibrium, the magnetic nanotube is expected to stabilize into a flux-terminating vortex state, in which all magnetization curls around the hollow core[8, 9]. Furthermore, in the nanotube structure, the inner void avoids the formation of Bloch points, facilitating the domain wall-based reversal process and leading to ultrafast information propagation through DWs[10].

Theoretical investigations in magnetic nanotube suggest various stable configurations at the remanent state, including a global vortex state, an onion state with parallel domains connected by 180° head-to-head and tail-to-tail domain walls, a uniform state where magnetization aligns along the tube's long axis and a mixture between these states[11, 12]. The existence and relative chirality (sense of rotation) of these states depend on the applied magnetic field history and nanotube dimensions (e.g., diameter, aspect ratio, wall thickness, and angle of applied magnetic fields)[7]. Previous attempts to experimentally characterize nanotube magnetization focused on mapping the stray field generated by bunches[13] or individual tubes[10]. For instance, Li et al., detected nearly vanishing contrast due to small remnant magnetization in a single nanotube through magnetic force microscopy attributing it to the vortex ground state[14]. Here, the azimuthally aligned vortex dominates the magnetic states of the nanotubes. Recent efforts utilized XMCD-PEEM to detect magnetization configurations with varying dimensions, but due to lack of high magnetic field, the saturation of nanotubes and subsequent detection of remanent magnetization state was not possible[15-18]. The Scanning SQUID was recently used to map stray fields produced by nanotubes as a function of position and applied field[10, 19]. Anisotropic magnetoresistance measurements on a single nanotube provided insights into magnetization reversal processes but did not yield conclusive inferences regarding magnetic states[20-24]. In summary, despite the availability of imaging techniques employing exotic instruments, characterizing the magnetic configuration of nanotubes using simple methodology remains elusive in both scope and detail.

In this manuscript, we have attempted to elucidate the magnetic state details using Remanence Field Ferromagnetic Resonance Spectroscopy (RFMR) for arrays of electrodeposited nanotubes. RFMR operates by applying a magnetization state preparation field and subsequently removing it, followed by measuring the zero-field resonance spectra in frequency scan mode. Realizing the fine magnetic state details of the nanotube arrays, are major advantage of RFMR. Initially, we applied an external field sufficient to saturate the nanotube arrays both perpendicular ($\theta_H = 90^o$) and parallel ($\theta_H = 0^o$) to the nanotube axis (see inset Figure 1(a) for $\theta_H$). The saturating field was then reduced to zero in these configurations, and RFMR spectra were obtained. Observations revealed opposite RFMR spectra, denoting opposite magnetic spin configurations after removing the positive and negative saturating fields for both $\theta_H = 0^o$ & $90^o$ configurations. However, RFMR spectra differed between the $\theta_H = 0^o$ and $\theta_H = 90^o$ cases. We observed a mixture of the non-uniform curling states with end vortex state (onion like curling state) at the end of the nanotubes for the $\theta_H = 0^o$ ($90^o$) and uniform magnetization states in the middle of the nanotubes for the $\theta_H = 0^o$ configuration. Additionally, the occurrence of opposite spin configuration and the presence of mixed ground magnetic state configurations in both $\theta_H = 0^o$ & $90^o$ cases were supported by micromagnetic simulations. Furthermore, different magnetic states were prepared by applying $H_{bias}$ after removing positive saturation field for $\theta_H = 0^o$ case, followed by frequency swept FMR. Here, $H_{bias}$ was limited to a lesser value as the system evolved through a range of magnetic states, each stable in a narrow field range. Through these FMR spectroscopy experiments, we investigated the evolution of the magnetization configuration in presence of a bias field for the magnetic nanotube arrays. These findings were further corroborated with First Order Reversal Curve measurements (FORC).

## Materials and Methods

$Ni_{30}Co_{70}$ nanotubes (NTs) were fabricated through electroless plating within ion track-etched polycarbonate membranes to ensure uniform tube lengths. These NTs, approximately 20 μm long and matching the thickness of the polycarbonate membranes, were, spaced 80 nm apart, with an aspect ratio (outer diameter (D)/ length (L)) 0.023 and density $10^8$ tubes-cm$^{-2}$. The fabricated NTs holds a diameter of ~ 470 nm and a wall thickness ~47 nm. Details of the synthesis and structural analysis have been documented elsewhere[25]. For the FMR measurements, a home-built, fully automated FMR setup was employed, utilizing a lock-in detection (EG&G 7265) technique to capture derivative spectra. The NTs array was placed on the coplanar waveguide which was further placed inside an electromagnet. Concurrently, a static DC magnetic field was applied perpendicular to the RF field generated by the waveguide. Field modulation was generated via a pair of Helmholtz coils at a reference frequency of 211.31 Hz. Frequency sweep FMR was conducted by sweeping the radio frequency until 20 GHz using the signal generator SMA100 B at a fixed or zero DC magnetic field. A Lakeshore VSM (7404) was utilized for the measurements of FORCs and the standard hysteresis loops. To comprehend the equilibrium spin configuration when the applied field aligns along $\theta_H = 0^o$ & $90^o$, we utilized the MuMax3 software package, employing the Landau-Lifshitz-Gilbert (LLG) equation formalism with finite difference discretization to understand the equilibrium magnetic configuration[26]. The magnetization dynamics is given by the LLG equation as:

$$\frac{d\hat{m}}{dt} = -\gamma \hat{m} \times \vec{H}_{eff} + \alpha \hat{m} \times \frac{d\hat{m}}{dt}$$

(1)

where $\gamma$ is the gyromagnetic ratio, $\hat{m}$ is the unit magnetization vector, $\alpha$ is the Gilbert damping coefficient and $\vec{H}_{eff}$ is the effective magnetic field given. In micromagnetic simulation initially the magnetization in the nanotubes, are considered to be random in nature. Further, a saturating magnetic field of 2T was applied along the nanotube axis (z axis, see Figure 1(a) inset) for $\theta_H = 0^o$ and perpendicular to the nanotube axis (x axis, see Figure 1(a) inset) for $\theta_H = 90^o$. In the saturated state, all the spins are aligned in the direction of the applied field. Subsequently, the magnetic fields were reduced to zero and the equilibrium magnetic states were obtained after energy minimization. We have used "ParaView"[27] to do post-processing of the magnetization profile. To model these extensive NT arrays, we employed a matrix of 3×3 NT arrays with a reduced length of 1000 nm (Due to the computational limit the length of 1 μm was taken while keeping the form factor $\beta$ (inner diameter/ outer diameter) and inter tube distance similar as of experimentally grown NTs. Also, the magnetic states for long nanotubes depend on $\beta$[15]). The simulation parameters used were as follows: saturation magnetization $M_S = 7.6 \times 10^5 \, Am^{-1}$, outer diameter = 470 nm, thickness = 47 nm, uniform exchange stiffness constant $A = 10 \, pJm^{-1}$ and damping constant $\alpha = 0.05$. Based on these parameters, a cubic cell size of 5 nm was chosen, smaller than the exchange length.

## Results & Discussion

Before analysing the RFMR spectra of the $Ni_{30}Co_{70}$ nanotubes (NTs) array, it's crucial to understand the equilibrium spin configuration when the applied field aligns along $\theta_H = 0^o \, \& \, 90^o$. For this purpose, we have utilized micromagnetic simulations. The magnetic parameters required for the same was obtained from standard hysteresis measurements conducted with the field applied along $\theta_H = 0^o \, \& \, 90^o$, depicted in Figure 1(a) and 1(b) respectively. The simulated hysteresis loops for $\theta_H = 0^o \, \& \, 90^o$ are presented in Figure 1(a) and 1(b) respectively, represented by red coloured dots, and show excellent agreement with the experimentally obtained loops. Our findings reveal that $\theta_H = 0^o \, (90^o)$ represents the easy (hard) axis for the NT arrays.

The simulated magnetization profile at zero field starting from positive and negative saturation for the $\theta_H = 0°$ configurations are displayed in Figure 1(c) and 1(d) respectively. It's worth noting that we considered the magnetization profile of the middle nanotube in a 3×3 matrix of NT arrays to illustrate the dipolar field effects from the nearest neighbouring nanotubes. In this configuration, the magnetization at the central part of the nanotube aligns along its long axis due to the strong magnetostatic field and two non-uniform curling end state with edge vortex state due to the decrease in magnetostatic energy in order to reduce the surface magnetic charge at the end of the nanotube. The magnetization chirality at the end of the nanotubes were found to be anticlockwise ($C_\uparrow$) and clockwise ($C_\downarrow$) in nature. Interestingly, depending on whether the preparation field starts from positive or negative saturation, the chirality reverses for the edge vortex state i.e. $C_\uparrow C_\downarrow$ becomes $C_\downarrow C_\uparrow$ state with the opposite magnetization direction for uniform magnetization state, as shown in Figure 1(c) and 1(d) respectively. In contrast to the $\theta_H = 0°$ configuration, the magnetization profile for the $\theta_H = 90°$ scenario displays a mixture of the "onion state" at the nanotube edges and the "curling state" at the centre at zero field, presented in Figure 1(e) and 1(f) respectively. Here Onion state, which defines parallel domains connected by 180° head-to-head and tail-to-tail domain walls and curling state is a pattern of magnetization reversal where the magnetic moments rotate in a spiralling fashion, with the extent of rotation increasing as the size of nanotube grows. The chirality of the onion state changes for this configuration as well depending on the application of positive or negative saturation field and its subsequent removal.

Next, we present the dynamic measurement at the remanent field starting from positive and negative saturation for the $\theta_H = 0°$ & $90°$ configuration, and the FMR absorption spectrum is depicted in Figure 2(a) and (b) respectively. Figure 2 exhibits several intriguing characteristics such as 1. Occurrence of the three (two) FMR modes for $\theta_H = 0°$ ($90°$), 2. The presence of two mirror symmetric peaks for $\theta_H = 0°$ and one for $\theta_H = 90°$ when the field is brought to zero field from positive and negative saturation and 3. One same peak in high frequency for $\theta_H = 0$ & $90°$ when the field is brought to zero field from positive and negative saturation.

The imaginary part of FMR absorption spectrum were fitted in frequency domain using a central difference quotient model given by[28]

$$\frac{d\chi}{d\omega} = Ai\omega \frac{\chi(\omega+\Delta\omega_s)-\chi(\omega-\Delta\omega_s)}{2\Delta\omega_s} + C \qquad (2)$$

where $C$ corrects the offset, A is the amplitude, $\Delta\omega_s$ is the frequency step. The magnetic susceptibility is given by[28],

$$\chi(\omega, H_{bias}) = \frac{\omega_M(\gamma\mu_0 H_{bias}+i\Delta\omega)}{(\omega_r(H_{bias}))^2-\omega^2+i\omega\Delta\omega} \qquad (3)$$

where $\omega_r$ is the resonance frequency, $\Delta\omega$ is the full width at half maximum linewidth and $\omega_M = \gamma\mu_0 M_s$. The frequency swept FMR spectra depicted in Figure 2(a) and (b) were fitted using the Eq$^n$ (2) and for $\theta_H = 0^o$ (90$^o$), three (two) FMR mode has been detected at 4.4 (5) GHz, 8.2 GHz and 15.9 (16.3) GHz, respectively. At zero magnetic field, the Kittel equation[29, 30] can be simplified to

$$f_0 = \frac{\mu_0 \gamma}{2\pi} \sqrt{\left(H_{eff} + (N_Y - N_{H\parallel}).M_S\right).\left(H_{eff} + (N_{H\perp} - N_{H\parallel}).M_S\right)}$$

(4)

where $H_{eff}$ is the effective field contains anisotropy field, demagnetization field, exchange field and dipolar field from the surrounding nanotubes, $N_{H\parallel}$, $N_{H\perp}$ local demagnetization factors along and perpendicular to the field, $N_Y$ demagnetization factor along the Y-axis, $\gamma$ the gyromagnetic ratio, $\mu_0$ the magnetic permeability of the free space, and $M_S$ the saturation magnetization. The edge vortex state has lesser $H_d$ in comparison to the curling and uniform magnetization state due to flux closure spin configuration[12]. Thus, it is evident from Eq$^n$ (4) that the FMR absorption corresponding to the edge vortex must occur at lower frequency in comparison to the curling and uniform magnetization mode. Thus, the FMR mode occurring at 4.4 GHz in Figure 2(a) correspond to the edge vortex state. Similarly, Onion state for $\theta_H = 90^o$, represent FMR mode at ~ 5 GHz. Additionally, depending on the choice of starting field for state preparation for instance positive or negative saturation field, the chirality of the vortex, and onion state for $\theta_H = 0^o$ & 90$^o$ configuration changes, observed through the mirror-symmetric FMR spectrum for both modes in Figure 2(a) and (b), respectively. Now we have two additional peaks in Figure 2(a) which might represent the uniform magnetization FMR mode or curling magnetization mode. Generally, in the curling mode, the magnetization is organized in a spiral or helical pattern, when the field is reversed from positive to negative saturation, the dynamic response of this curling state to the changing field remains consistent, it is due to similar sense of rotation while field brought to positive or negative saturation (figure 1(c & d)), which is why the FMR peak remains the same. The FMR peak related to curling magnetization mode lies in higher frequency as compared to uniform magnetization mode as the energy of curling mode is higher than the uniform mode[31]. This indicates that the dynamic response of curling state lies in higher frequency and is consistent to magnetic field.

Moreover, in a general scenario, the uniform magnetization state shows opposite FMR peaks as a consequence of polarity reversal of the saturation field. Further, a general trend of uniform magnetization mode is a continuous shift in resonance field with the microwave frequency. Both features of the uniform mode can be readily confirmed through frequency sweep FMR measurements conducted with a NiFe (10nm) thin film (see supplementary information). Notably, among the remaining two peaks (second and third peak) in Fig. 2(a), second peak exhibits opposite FMR spectra upon the polarity reversal of saturation field. Therefore, it is speculated that the second peak in Figure 2(a) corresponds to the uniform magnetization mode. To confirm a continuous shift in resonance peak with microwave frequency as the biasing field increase, frequency-sweep FMR measurements were performed in presence of biasing fields ($H_{bias}$). These measurements were taken at various bias field strengths ranging from 3000 Oe to 0 Oe and RFMR spectra for various field for $\theta_H = 0^o$ shown in Figure 3(a). In Figure 3(b), the dispersion of different FMR peak frequencies with magnetic field is depicted (see supplement for the fitting of the frequency swept FMR absorption spectrum using Eq$^n$ (2)). It is worth noting that during this measurement, following the application of a positive saturation field, the field was reduced to $H_{bias}$. The speculated uniform mode shows an expected increment in resonance frequency as the biasing field increases, following an initial negative dispersion (Fig. 3(b)). To understand the negative dispersion of the field, one must consider invoking the Kittel dispersion $f_{res} \propto \sqrt{H_I(H_I + M)}$ where, the internal field $H_I = H_d + H_{bias}$. With increasing $H_{bias}$, the negative dispersion correlates that the magnetization direction is opposite to the applied field and $|H_d| > H_{bias}$. However, with a further increase in $H_{bias}$, the spins get aligned along the $H_{bias}$, one can envision a scenario where the magnetization configuration in one part of the nanotube nucleates. Therefore, the second peak in figure 2(a) represents the uniform magnetization mode. Further, based on our RFMR findings, we endeavoured to detect the magnetization configuration in the presence of a bias fields. We noticed some interesting points in Fig. 3(a). The amplitude of the FMR absorption spectra for the uniform mode goes up as the bias field ($H_{bias}$) increases, but the peak position of the high-frequency mode stays the same. The high-frequency modes corresponding to the curling magnetization state is found to occur for both $\theta_H = 0^o$ & $90^o$ configuration at 15.9 GHz, 16.3 GHz as depicted in Figure 2(a) and Figure 2(b) (see Figure 2(b) inset), respectively remain in place with the bias field, as shown in Figure 3(b). Furthermore, the curling configuration at $\theta_H = 0^o$ experiences negligible torque from $H_{RF}$ compared to the $\theta_H = 90^o$, leading to smaller absorption and peak intensity in the FMR spectrum of Figure 2(b) compared to Figure 2(a) (see supplement). Moreover, the intensification of the FMR absorption spectrum with increasing $H_{bias}$ suggests the evolution of uniform magnetization as compared to curling state in the entire region of the nanotubes. Hence, with increasing $H_{bias}$, nucleation is likely to occur within the curling state. Based on these findings, it can be inferred that nucleation within the curling state occurs within the field range of 500 to 600 Oe. Additionally, the increase in $f_{res}$ with increasing $H_{bias}$ indicates the spin configuration aligning with the applied field direction where $H_I > 0$. Moreover, the disappearance of the lower frequency mode signifies a complete change in the nature of the edge vortex spin configuration.

To confirm the evolution of the edge vortex mode, field sweep FMR measurements were conducted. During these measurements, the application of a field sweep evolves the magnetic states in a nanotube array, and their absorption spectra were recorded at a fixed microwave frequency. Field sweep FMR results shown in figure 4(a) reveals that the vanishing of the peak (encircled) beyond 5 GHz microwave frequencies suggests the presence of the edge vortex state within the nanotube in lower frequency regime. Further, this peak evolved into new FMR peak at 6 GHz, the new evolved FMR peak (around 500 Oe) corresponds to the curling state. We have observed similar features in our micromagnetic simulations. The simulated magnetization profile of the nanotube in presence of $H_{bias} = 200\ Oe\ \&\ 500\ Oe$ is shown in Figure 4(c). Here, the nucleation of curling state (bottom) is shown from the edge vortex state (top) as the field evolve from 200 Oe to 500 Oe. Further, the variation of $M_z$ vs. length of nanotube at different $H_{bias}$ is exhibited in Figure 4(d) which shows the overall magnetic states present at a fixed field. One can clearly observe that vortex state (average magnetization zero) at the edge of nanotube and central part of nanotube have uniform magnetization either +1 or -1 from figure 4(d). (See supplement for the evolution of the magnetic spin configuration with different $H_{bias}$). The curling state lies in between portion of vortex and uniform state. The nucleation in curling state occurs at $H_{bias}$= 400 Oe and with increasing $H_{bias}$ the volume of uniform magnetization state expands as observed micromagnetically, which is supporting the findings from RFMR and subsequent FMR spectra in the preceding section. Discrepancies between the nucleation field values obtained from FMR spectra and micromagnetic simulations can be attributed to nanotube magnetization experiencing pinning forces from defects, unaccounted for in the simulations.

To further verify nucleation field and understand magnetization reversal in nanotubes at $\theta_H = 0^o$, we employed the First Order Reversal Curve (FORC) method (see supplement). Figure 5(a) illustrates the FORC diagram in $H_a - H_b$ coordinates (see supplement for the details of the FORC) with the corresponding color scale kept as an inset as a measure of the FORC distribution ($\rho$). The maximum weight of the $\rho$ is indicated by velvet color while the minimum is shown in blue color. The FORC distribution is becoming non-zero near ~ 400 Oe (see line 1 in Figure 5(a)) indicates the onset of irreversible processes. This irreversibility projects nucleation of states, aligning with our earlier FMR and micromagnetic simulation-based observations. The irreversible process peaks around a reversal field value of ~ -50 Oe (line 2 in Figure 5(a)). For an understanding of interaction field distribution in nanotubes at $\theta_H = 0^o$, a statistical analysis was performed. Figure 5(b) represents the interaction field distribution at field values indicated by blue lines through FORC maxima in $H_u - H_c$ coordinates (see supplement). Notably, the maximum FORC distribution centres around $H_u$= 8 Oe. According to Pike et al., in assemblies of non-interacting single-domain particles, introduction of a mean interacting field results in the displacement of the peak of FORC distribution depending on the nature of the mean field interaction[32, 33]. Typically, positive displacement of the $H_u=0$ axis refers dipolar interaction whereas negative displacement corresponds to exchange like interaction[34]. In our nanotubes, at $\theta_H = 0^o$ configuration, the mean interacting field appears dipolar in nature, further corroborating with our FMR findings.

Now, we have a picture of the configuration of magnetic states corresponding to all the FMR peaks within FMR spectrum which is corroborating with micromagnetic simulations and FORC. In our frequency sweep experiments, we observed that the peak found at lower frequencies corresponds to the features of the edge vortex state for $\theta_H = 0^o$, disappearing similarly beyond exposure of 500 Oe field and the nucleation occurs at the same field. By employing the Kittel formula, we determined the resonance frequency linked with the vortex state within the lower frequency range and uniform magnetization mode follows the opposite FMR spectra while field brought to positive and negative saturation and continuous shift with microwave frequency. The curling mode lies in higher frequency regime and having similar FMR peaks. This confirms that FMR can detect the remanent spin configuration of a system.

# Conclusion

In conclusion, we've elucidated the magnetic configuration of nanotubes through RFMR and subsequent FMR absorption at varying bias fields. Presently, characterization of nanotubes spin configurations is unavailable through net magnetization measurements. This characterization largely exists theoretically through micromagnetic simulations due to limitations in exotic measurement techniques such as MFM, PEEM and XMCD. Our approach, utilizing both frequency sweep and field sweep FMR measurements relying solely on RFMR, offers an elegant solution. Building on RFMR information coupled with remnant micromagnetic picture, FMR absorption spectra obtained at different bias fields allowed characterization of magnetization states which remain elusive through simple magnetic measurements technique and on the other hand at fixed frequency the field sweep FMR measurement technique coupled with evolution of the magnetic states. While our demonstration focuses on nanotubes, we believe RFMR holds promise across various interacting nanomagnetic systems, particularly in the expanding realm of 3D artificial spin systems where imaging poses inherent challenges. Moreover, RFMR serves as an appealing state readout solution for recent neuromorphic and wave computation schemes, leveraging extensive spaces for next-generation computing.

## ASSOCIATED CONTENT

**Supporting Information**: Details of field sweep FMR, fitting of frequency swept FMR peaks, Interaction of RF field with the spin configuration, Micromagnetic simulation at different $H_{bias}$ field and First Order Reversal Curve measurement.

## AUTHOR INFORMATION

### Corresponding Author

**Debangsu Roy**- *Department of Physics, Indian Institute of Technology Ropar, Rupnagar 140001, India;*

Email: debangsu@iitrpr.ac.in

### Authors

**Abhishek Kumar**- *Department of Physics, Indian Institute of Technology Ropar, Rupnagar 140001, India*

https://orcid.org/0000-0001-6440-0490

**Chirag Kalouni**- *Department of Physics, Indian Institute of Technology Ropar, Rupnagar 140001, India*

**Raghavendra Posti**- *Department of Physics, Indian Institute of Technology Ropar, Rupnagar 140001, India; orcid.org/0000-0003-1632-0095 23*

**Vivek K Malik-** *Department of Physics, Indian Institute of Technology Roorkee, Roorkee 247667, India*

**Dhananjay Tiwari**- *Silicon Austria Labs GmbH, Sandgasse 34, Graz 8010, Austria*

## ACKNOWLEDGMENTS

DR acknowledges the financial support from the Department of Atomic Energy (DAE) under project no. 58/20/10/2020-BRNS/37125 & Science and Engineering Research Board (SERB) under project no. CRG/2020/005306. AK acknowledges the financial assistance from UGC.

# References

1. A. Torti, V. Mondiali, A. Cattoni, M. Donolato, E. Albisetti, A. M. Haghiri-Gosnet, P. Vavassori and R. Bertacco, *Applied Physics Letters*, 2012, **101**, 142405.
2. A. Sebastian, M. Le Gallo, R. Khaddam-Aljameh and E. Eleftheriou, *Nature Nanotechnology*, 2020, **15**, 529-544.
3. A. Hoffmann and S. D. Bader, *Physical Review Applied*, 2015, **4**, 047001.
4. P. Landeros and Á. S. Núñez, *Journal of Applied Physics*, 2010, **108**, 033917.
5. G. Venkat, D. A. Allwood and T. J. Hayward, *Journal of Physics D: Applied Physics*, 2024, **57**, 063001.
6. M. Yan, C. Andreas, A. Kákay, F. García-Sánchez and R. Hertel, *Applied Physics Letters*, 2012, **100**, 252401.
7. R. Streubel, P. Fischer, F. Kronast, V. P. Kravchuk, D. D. Sheka, Y. Gaididei, O. G. Schmidt and D. Makarov, *Journal of Physics D: Applied Physics*, 2016, **49**, 363001.
8. J. A. Otálora, J. A. López-López, P. Landeros, P. Vargas and A. S. Núñez, *Journal of Magnetism and Magnetic Materials*, 2013, **341**, 86-92.
9. P. Landeros, S. Allende, J. Escrig, E. Salcedo, D. Altbir and E. E. Vogel, *Applied Physics Letters*, 2007, **90**, 102501.
10. A. Buchter, J. Nagel, D. Rüffer, F. Xue, D. P. Weber, O. F. Kieler, T. Weimann, J. Kohlmann, A. B. Zorin, E. Russo-Averchi, R. Huber, P. Berberich, A. Fontcuberta i Morral, M. Kemmler, R. Kleiner, D. Koelle, D. Grundler and M. Poggio, *Physical Review Letters*, 2013, **111**, 067202.
11. A. P. Chen, J. Gonzalez and K. Y. Guslienko, *Materials Research Express*, 2015, **2**, 126103.
12. C. Sun and V. L. Pokrovsky, *Journal of Magnetism and Magnetic Materials*, 2014, **355**, 121-130.
13. J. Escrig, S. Allende, D. Altbir and M. Bahiana, *Applied Physics Letters*, 2008, **93**, 023101.
14. D. Li, R. S. Thompson, G. Bergmann and J. G. Lu, *Advanced Materials*, 2008, **20**, 4575-4578.
15. P. Landeros, O. J. Suarez, A. Cuchillo and P. Vargas, *Physical Review B*, 2009, **79**, 024404.
16. M. Wyss, A. Mehlin, B. Gross, A. Buchter, A. Farhan, M. Buzzi, A. Kleibert, G. Tütüncüoglu, F. Heimbach, A. Fontcuberta i Morral, D. Grundler and M. Poggio, *Physical Review B*, 2017, **96**, 024423.
17. R. Streubel, L. Han, F. Kronast, A. A. Ünal, O. G. Schmidt and D. Makarov, *Nano Letters*, 2014, **14**, 3981-3986.
18. J. Hurst, A. De Riz, M. Staňo, J.-C. Toussaint, O. Fruchart and D. Gusakova, *Physical Review B*, 2021, **103**, 024434.
19. D. Vasyukov, L. Ceccarelli, M. Wyss, B. Gross, A. Schwarb, A. Mehlin, N. Rossi, G. Tütüncüoglu, F. Heimbach, R. R. Zamani, A. Kovács, A. Fontcuberta i Morral, D. Grundler and M. Poggio, *Nano Letters*, 2018, **18**, 964-970.
20. D. Rüffer, R. Huber, P. Berberich, S. Albert, E. Russo-Averchi, M. Heiss, J. Arbiol, A. Fontcuberta i Morral and D. Grundler, *Nanoscale*, 2012, **4**, 4989-4995.
21. M. Zimmermann, T. N. G. Meier, F. Dirnberger, A. Kákay, M. Decker, S. Wintz, S. Finizio, E. Josten, J. Raabe, M. Kronseder, D. Bougeard, J. Lindner and C. H. Back, *Nano Letters*, 2018, **18**, 2828-2834.
22. K. Baumgaertl, F. Heimbach, S. Maendl, D. Rueffer, A. Fontcuberta i Morral and D. Grundler, *Applied Physics Letters*, 2016, **108**, 132408.
23. A.-P. Chen, J. M. Gonzalez and K. Y. Guslienko, *Journal of Applied Physics*, 2011, **109**, 073923.
24. R. Sharif, S. Shamaila, M. Ma, L. D. Yao, R. C. Yu, X. F. Han and M. Khaleeq-ur-Rahman, *Applied Physics Letters*, 2008, **92**, 032505.
25. D. Tiwari, M. C. Scheuerlein, M. Jaber, E. Gautier, L. Vila, J.-P. Attané, M. Schöbitz, A. Masseboeuf, T. Hellmann, J. P. Hofmann, W. Ensinger and O. Fruchart, *Journal of Magnetism and Magnetic Materials*, 2023, **575**, 170715.

26. A. Vansteenkiste, J. Leliaert, M. Dvornik, M. Helsen, F. Garcia-Sanchez and B. Van Waeyenberge, *AIP Advances*, 2014, **4**, 107133.
27. J. P. Ahrens, B. Geveci and C. C. Law, 2005.
28. H. Maier-Flaig, S. T. B. Goennenwein, R. Ohshima, M. Shiraishi, R. Gross, H. Huebl and M. Weiler, *Review of Scientific Instruments*, 2018, **89**, 076101.
29. C. Kittel, *Physical Review*, 1948, **73**, 155-161.
30. A. Vanstone, J. C. Gartside, K. D. Stenning, T. Dion, D. M. Arroo and W. R. Branford, *New Journal of Physics*, 2022, **24**, 043017.
31. C.-R. Chang, C. M. Lee and J.-S. Yang, *Physical Review B*, 1994, **50**, 6461-6464.
32. C. R. Pike, A. P. Roberts and K. L. Verosub, *Journal of Applied Physics*, 1999, **85**, 6660-6667.
33. R. Tanasa, C. Enachescu, A. Stancu, J. Linares, E. Codjovi, F. Varret and J. Haasnoot, *Physical Review B*, 2005, **71**, 014431.
34. D. Roy, K. V. Sreenivasulu and P. S. Anil Kumar, *Applied Physics Letters*, 2013, **103**, 222406.

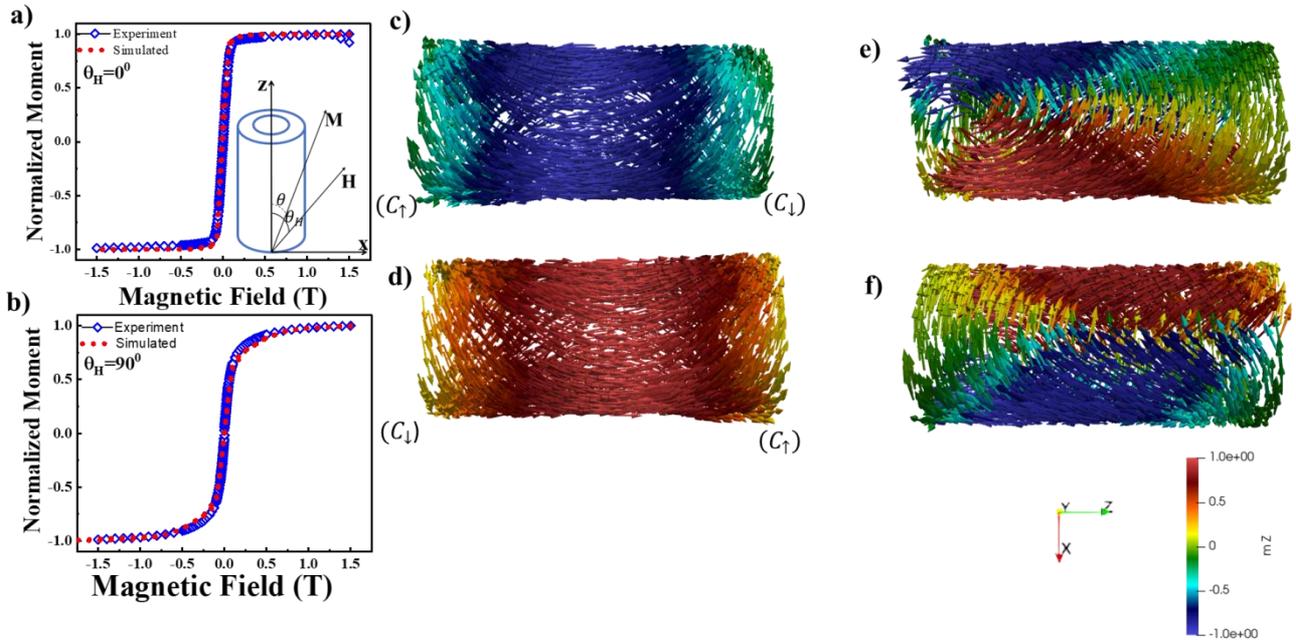

**Figure 1** Comparison of experimentally obtained *M-H* measurement and simulated *M-H* for **(a)** $\theta_H = 0^o$ case, (inset) geometrical representation of a nanotube where $\theta_H$ is the angle of the applied field with nanotube axis. **(b)** for $\theta_H = 90^o$ configuration. Micromagnetically obtained magnetization profile of nanotube **(c)** from positive saturation, **(d)** negative saturation to $H_{bias}$ =zero Oe for $\theta_H = 0^o$ case and **(e)** from positive saturation and **(f)** negative saturation to $H_{bias}$ =zero Oe for $\theta_H = 90^o$ case. The color scale denotes the z-component of magnetization where red (blue) color is for +1 (-1).

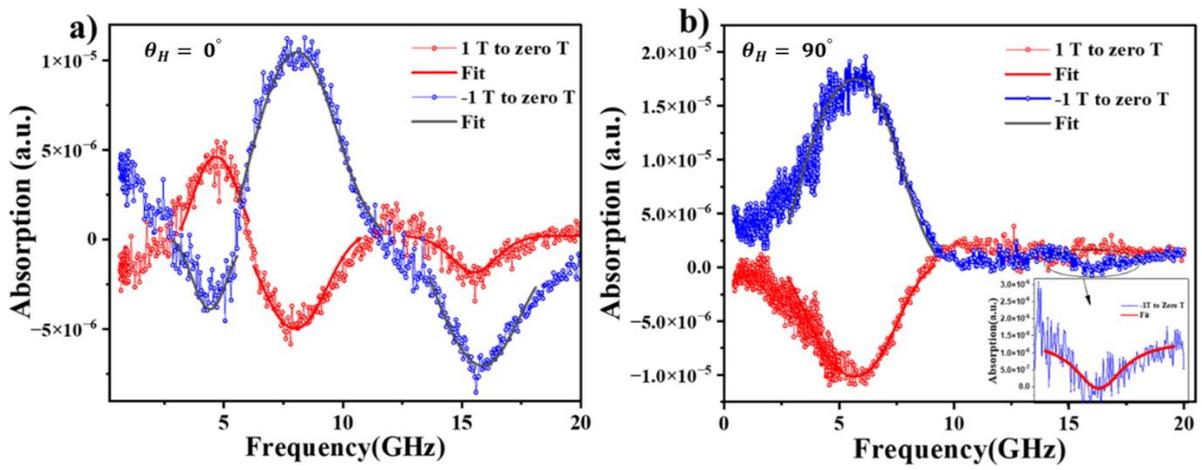

**Figure 2** Variation of the FMR absorption with frequency at remanent field (RFMR) for (a) $\theta_H = 0^o$ and (b) $\theta_H = 90^o$ showing the opposite magnetic state. The solid line denotes the fitting of the spectrum using Eq$^n$ (2). The inset in Figure 2(b) shows the expanded portion of the graph in Figure 2(b).

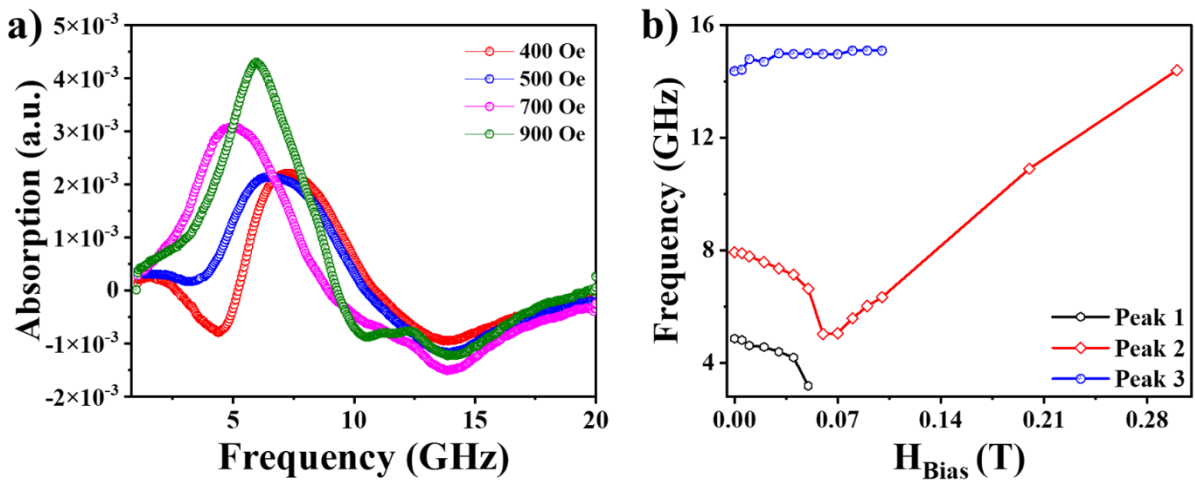

**Figure 3(a)** Frequency swept FMR absorption at different $H_{bias}$ field **(b)** Experimentally observed dispersion of the resonance frequency with different $H_{bias}$ (the solid lines are guide to the eye).

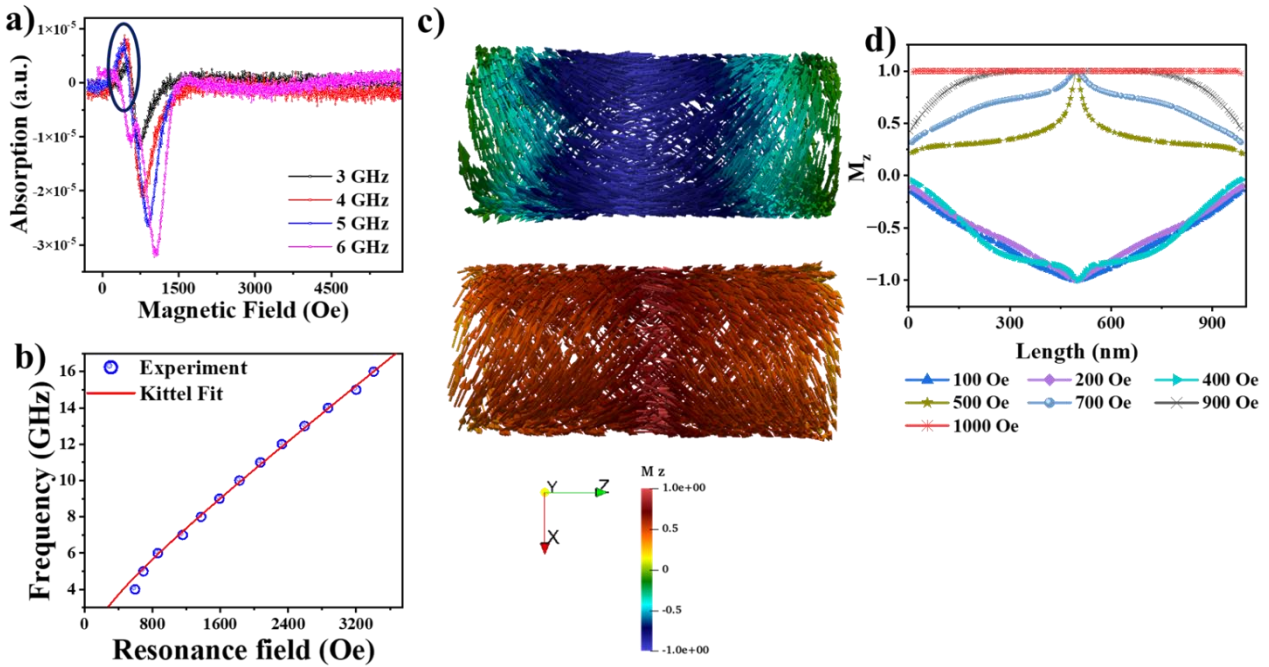

**Figure 4(a)** Field sweep FMR spectrum of Nanotube array at 3-6 GHz microwave frequencies, FMR peak representing the evolution of magnetic state is encircled and the maximum intense peak represent the uniform magnetization state **(b)** Resonance frequency versus magnetic resonance field corresponding to uniform magnetization state **(c)** Micromagnetically obtained magnetization profile at $H_{bias}$ = 200 & 500 Oe. The color scale denotes the z-component of magnetization where red (blue) color is for +1 (-1). **(d)** The spatial variation of $M_Z$ component (obtained Micromagnetically) of the magnetization with nanotube length at different $H_{bias}$.

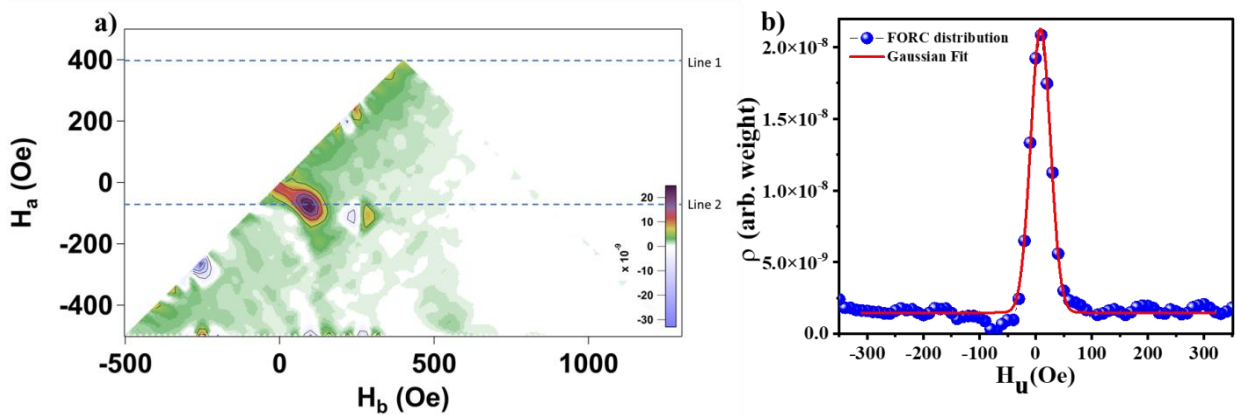

**Figure 5 (a)** FORC distributions for nanotube in $H_a$-$H_b$ co-ordinate. **(b)** Interaction field profile for nanotube. The red line corresponds to the Gaussian fitting.

# Supplementary Information

# Correlation of Magnetic State Configurations in Nanotubes with FMR spectrum

Abhishek Kumar[1]#, Chirag Kalouni[1]#, Raghvendra Posti[1], Vivek K Malik[2], Dhananjay Tiwari[3], and Debangsu Roy[1]*

[1]Department of Physics, Indian Institute of Technology Ropar, Rupnagar 140001, India

[2]Department of Physics, Indian Institute of Technology Roorkee, Roorkee 247667, India

[3]Silicon Austria Labs GmbH, Sandgasse 34, Graz 8010, Austria

## S1 Frequency swept FMR spectra of 10nm NiFe thin film:

FMR spectra of sputtered Si/SiO$_2$/NiFe(10nm) thin film at 25-watt DC power deposited at 2*10e-3 mbar Ar pressure was measured by using FMR spectroscopy. Initially the thin film was saturated by applying a 5000 Oe in-plane field and then field was reduced to $H_{Bias}$=10,20,30,40,50 Oe subsequently. The results suggested that the FMR peak corresponding to uniform FMR mode is shifted with the increase in $H_{Bias}$ shown in Figure S1 and it is obvious that the precessional frequency increase with the applied field. The next confirmation towards the FMR peak corresponding to uniform mode in our nanotube sample was observed when we took the spectra for various $H_{Bias}$ after saturating the film in positive and negative direction. The mirror symmetric FMR spectra observed shown in inset of Figure S1. Hence the second FMR peak in the figure 2(a) in main text represent the uniform magnetization mode.

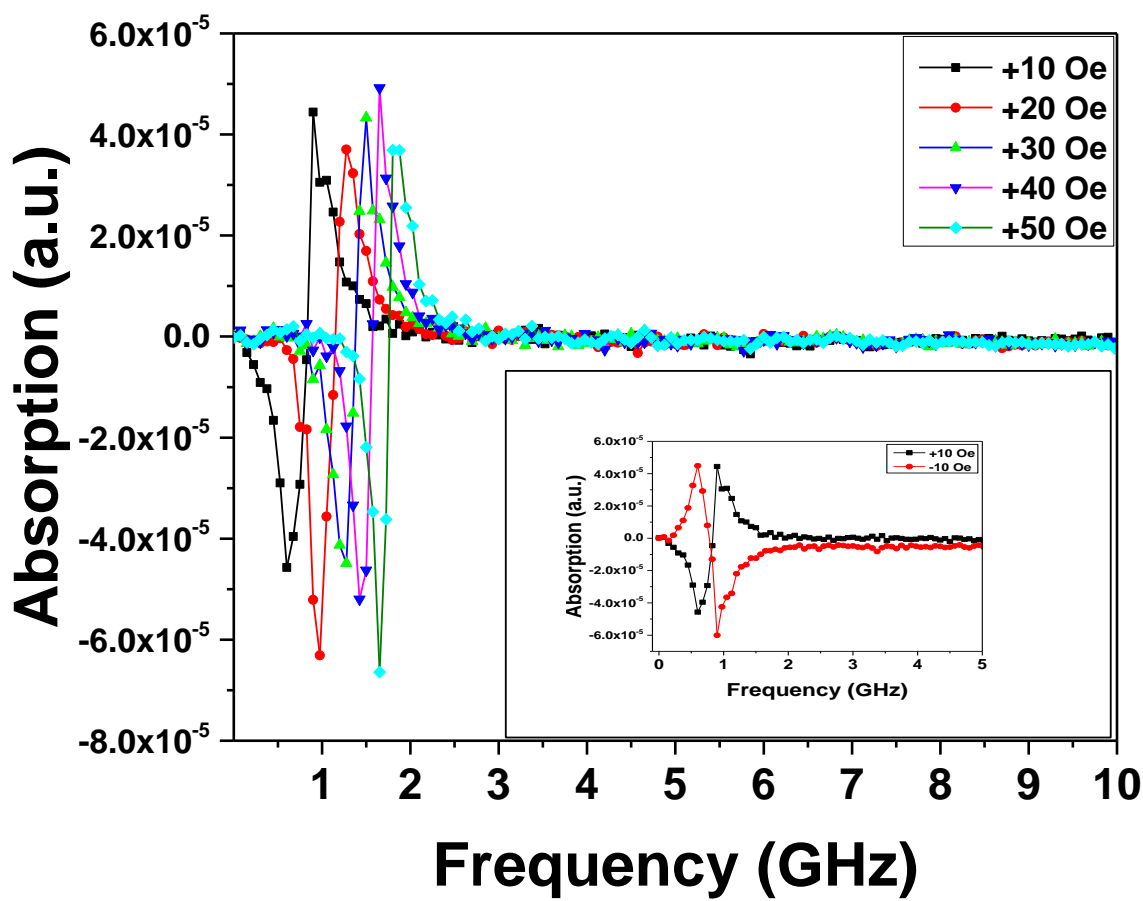

**Figure S1** *Frequency sweep FMR spectra for 10 nm NiFe thin film at different $H_{Bias}$ field after in-plane saturating the film, Mirror symmetric FMR spectra for uniform magnetization mode (inset)*

# S2 Fitting of FMR peaks and corresponding states:

The fitting of the frequency-swept FMR absorption spectra using Equation (1) from the main text is illustrated in Figure S2. In this case, the bias field strengths vary from 0 Oe to 3000 Oe for $\theta_H = 0°$. The fitted parameters are tabulated in Table 1. Here, $f_{r1}$ and $f_{r2}$ represent the resonance frequencies corresponding to the vortex state and uniform state, and $f_{r3}$ corresponds to the resonance frequency of the curling magnetization state.

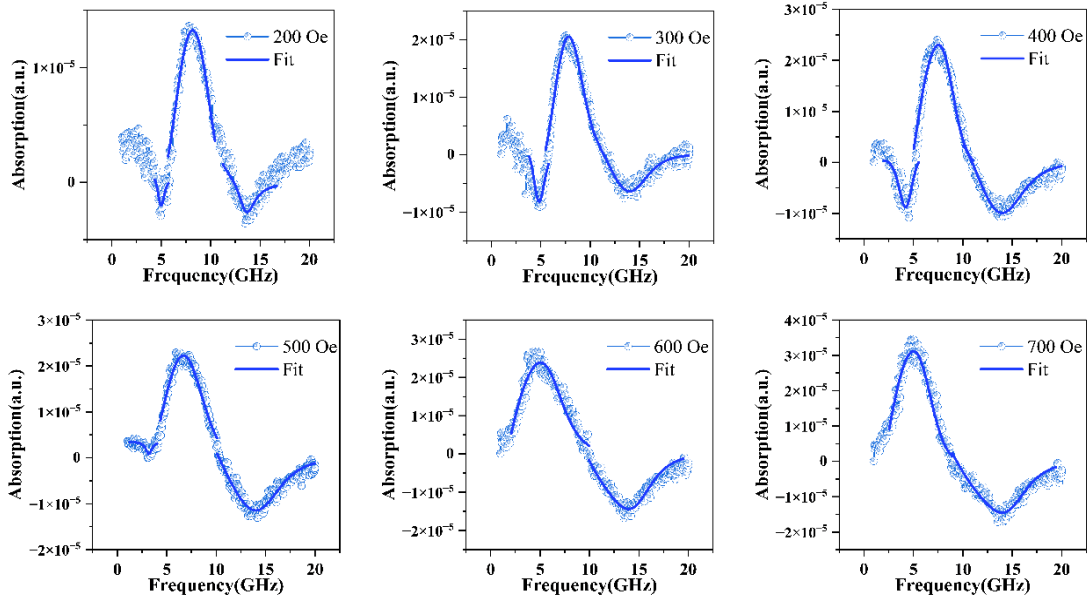

***Figure S2****. Fitted FMR peaks by using Eq. 2 (main text) at different $H_{bias}$ (200-700 Oe) for $\theta_H = 0°$.*

**Table 1. Resonance frequency obtained from fitting of the FMR spectra.**

| $H_{bias}$ | $f_{r1}$ (GHz) | $error_{fr1}$ (GHz) | $f_{r2}$ (GHz) | $error_{fr2}$ (GHz) | $f_{r3}$ (GHz) | $error_{fr3}$ (GHz) |
|---|---|---|---|---|---|---|
| 0 Oe | 4.86 | 0.012 | 7.93 | 0.015 | 14.37 | 0.012 |
| 50 Oe | 4.81 | 0.021 | 7.89 | 0.21 | 14.42 | 0.011 |
| 100 Oe | 4.61 | 0.031 | 7.78 | 0.03 | 14.8 | 0.035 |
| 200 Oe | 4.56 | 0.022 | 7.58 | 0.015 | 14.7 | 0.002 |
| 300 Oe | 4.39 | 0.02 | 7.35 | 0.009 | 15.0 | 0.002 |
| 400 Oe | 4.19 | 0.034 | 7.13 | 0.012 | 14.98 | 0.032 |
| 500 Oe | 3.18 | 0.02 | 6.63 | 0.029 | 15.0 | 0.021 |
| 600 Oe | | | 5.02 | 0.023 | 14.98 | 0.011 |
| 700 Oe | | | 5.05 | 0.021 | 14.97 | 0.010 |
| 800 Oe | | | 5.58 | 0.023 | 15.1 | 0.021 |
| 900 Oe | | | 6.02 | 0.003 | 15.1 | 0.025 |
| 1000 Oe | | | 6.33 | 0.01 | 15.1 | 0.03 |
| 2000 Oe | | | 10.9 | 0.01 | | |
| 3000 Oe | | | 14.4 | 0.006 | | |

# S3 Interaction of RF field with the spin configuration of the nanotubes.

The FMR peak amplitude depends on the interaction of the RF field ($H_{RF}$) with the spin configuration at the corresponding state. Figure S3(a) & S3(b) shows the interaction between $H_{RF}$ and spin configuration at $\theta_H = 0°$ & $\theta_H = 90°$.

At $\theta_H = 0°$, Figure 1 in the main text depicts the occurrence of the curling state with edge vortex state and uniform magnetization state. For the vortex state, $M_{total} = \sum M = 0$ hence $M_{total} \cdot H_{RF} = 0$. In the uniform magnetization state, the magnetization is mostly aligned along the $z$-axis, whereas $H_{RF}$ is along the $y$ axis. For the uniform magnetization state, $M_{total} \cdot H_{RF}$ approaches zero. However, for the curling state, $M_{total} = \sum M > 0$ and the $\angle M_{total}, H_{RF} \neq 90°$. Therefore, the curling state would correspond to a smaller amplitude of the FMR peak, whereas the uniform magnetization mode would lead to a maximum contribution.

At $\theta_H = 90°$, Figure 1(e & f) in the main text displays a mixture of the "onion state" at the nanotube edges and the "curling magnetization state" at the centre for the nanotube at zero applied field. For the onion state, $M_{total} = \sum M \neq 0$ hence $M_{total} \cdot H_{RF} \neq 0$., This results in a maximum amplitude of the FMR peak. However, in the curling state, where $M_{total} = \sum M > 0$ and the angle between $M_{total}$ and $H_{RF}$ is not equal to 90°, the curling state, along with the edge onion state, results in a relatively smaller amplitude of the FMR peak. In contrast, the curling magnetization mode contributes insignificantly.

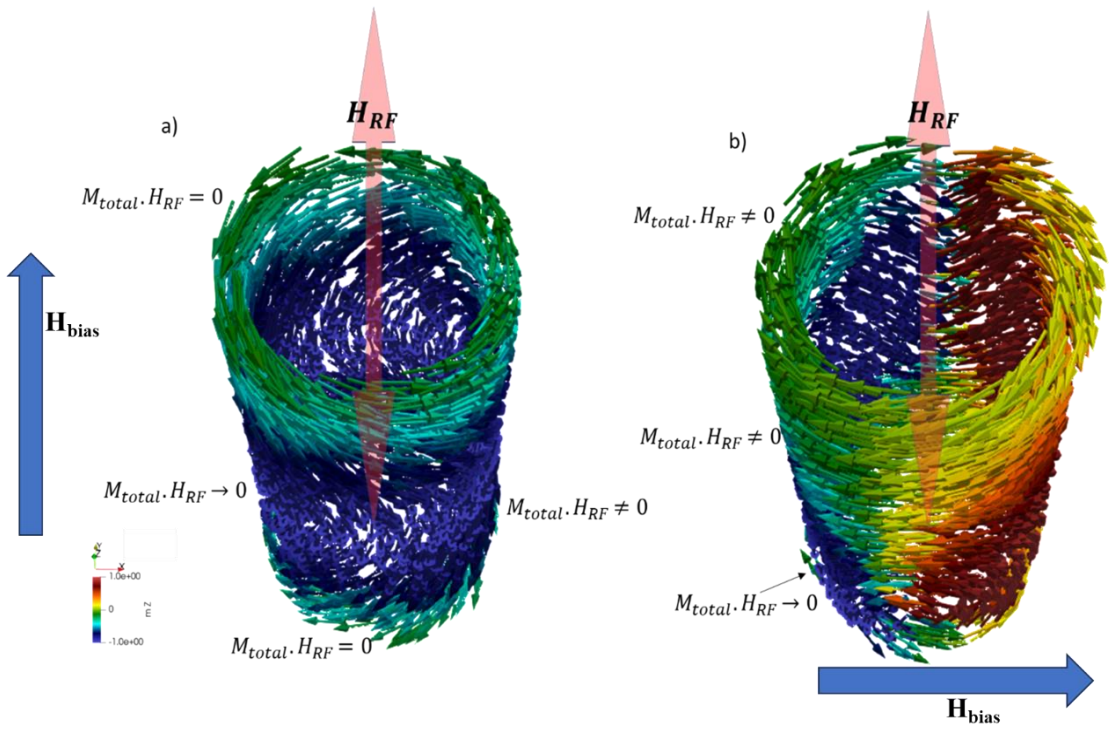

***Figure S3*** *(a) Interaction of the RF field with the curling with edge vortex and uniform magnetization state ($M_{total} \perp H_{RF}$) for $\theta_H = 0°$ (b) Interaction of the RF field with uniform magnetization state & onion like curling state for $\theta_H = 90°$. The color scale denotes the z-component of magnetization where red(blue) color is for +1(-1). The colored axis gives the idea of axis of magnetization profiles of NT.*

## S4 Micromagnetic simulation at different $H_{bias}$ field and evolution of magnetic states:

The micromagnetic results for different $H_{bias}$ field values at $\theta_H = 0^o$ are shown in Figure S4. During this simulation, we reduced the field to $H_{bias}$ after applying a positive saturation field. Through the spin configuration, we attempted to understand the evolution of the magnetization states. In the configuration $\theta_H = 0^o$, the magnetization at the central part of the nanotube aligns along its long axis due to the strong magnetostatic field and two non-uniform curling end state with edge vortex state due to the decrease in magnetostatic energy in order to reduce the surface magnetic charge at the end of the nanotube. The micromagnetic picture confirmed the presence of these states in the system within the $H_{bias}$ field range of 100–400 Oe (Figure S4). Further increment in the $H_{bias}$ from 400 Oe the spin states evolved and nucleation of the spins occur at 400-500 Oe bias field. However, with increasing $H_{bias}$ from 500 to 800 Oe, the spin configuration has changed from edge vortex type curling state to uniform magnetization state.

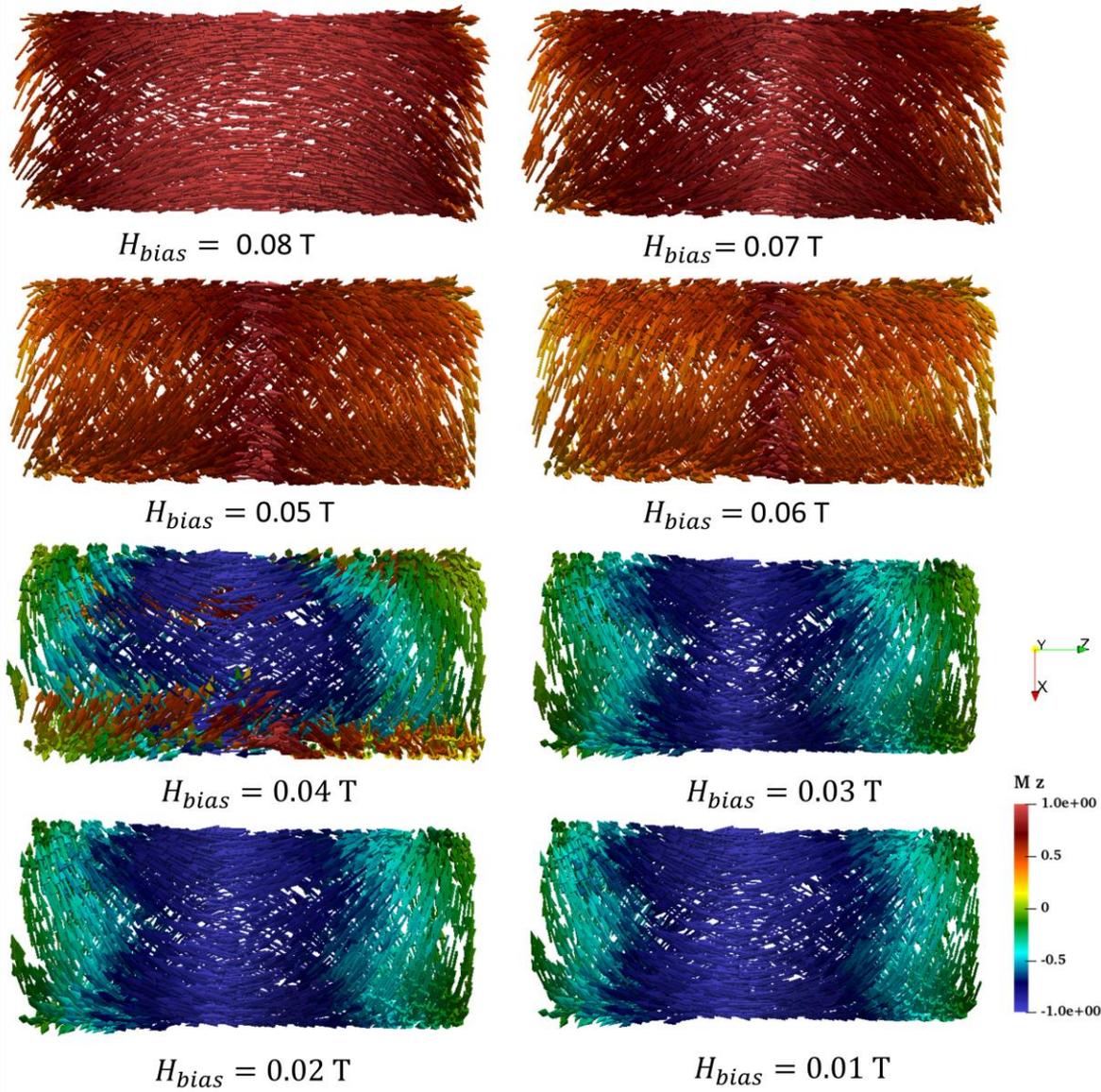

***Figure S4***. *Magnetic state evolution for $H_{bias}$=800 to 100 Oe field when field is applied parallel ($\theta_H = 0^o$) to the nanotube axis. The color scale denotes the z-component of magnetization where red(blue) color is for +1 (-1)*

## S5 First Order Reversal Curve (FORC) measurement for $\theta_H = 0^o$:

To investigate the magnetic states and comprehend the magnetization reversal in nanotubes with the applied field along $\theta_H = 0^o$, the First Order Reversal Curve (FORC) method has been employed. In this technique, a Lakeshore VSM (7404) was used for FORC measurements. Initially, the nanotubes were brought to positive saturation by applying a large magnetic field (1.5 Tesla). Subsequently, the magnetic field was reduced to $H_a$ (reversal field). From $H_a$ to positive saturation, the magnetization was measured, forming a single FORC. These measurements were conducted within the reversal field range of 700 Oe to -700 Oe, with a field step of 20 Oe. The magnetization on a FORC curve at an applied field $H_b$ for a reversal field of $H_a$ is denoted by M ($H_a$, $H_b$) where $H_b \geq H_a$. The FORC distribution can be calculated using the following formula

$$\rho(H_a, H_b) = -\frac{1}{2} \frac{\partial^2 M(H_a, H_b)}{\partial H_a \partial H_b}$$

from the data of consecutive measurement points on consecutive reversal curves. In this manuscript, we have used FORCinel based on the locally weighted regression smoothing algorithm (LOESS)[1,2] to obtain the FORC and related diagram from the raw data.

It is generally convenient to define a new set of coordinates $(H_u, H_c)$ instead of $(H_a, H_b)$. [$H_u = \frac{H_a + H_b}{2}$, $H_c = \frac{H_a - H_b}{2}$] to depict a FORC diagram. Both coordinate systems were employed to analyze the results. In our measurement the non-zero $\rho$ corresponds to the onset of the irreversible magnetization switching in the nanotubes occurring in the field range of ~ +400 Oe as observed in Figure 4(a). To understand the interaction mechanism in the nanotubes, we obtained the FORC distribution as a contour plot in $(H_u, H_c)$ coordinates, shown in Figure S5. Here, the maximum weight of the $\rho$ is indicated by velvet colour while the minimum is shown in blue colour. Generally, the projection of $\rho$ on to $H_u$ axis at a particular coercive field value can be visualized as the distribution of the interaction field strength. The interaction field profile is a measure of the exchange coupling or the dipolar interaction strength present in the nanotubes with the applied field along $\theta_H = 0°$. We obtained the interaction field profile at the coercive field values indicated by the blue lines through the maxima of the FORC distribution in Figure S5. The interaction field distribution is shown in Figure 5(b) of the main text.

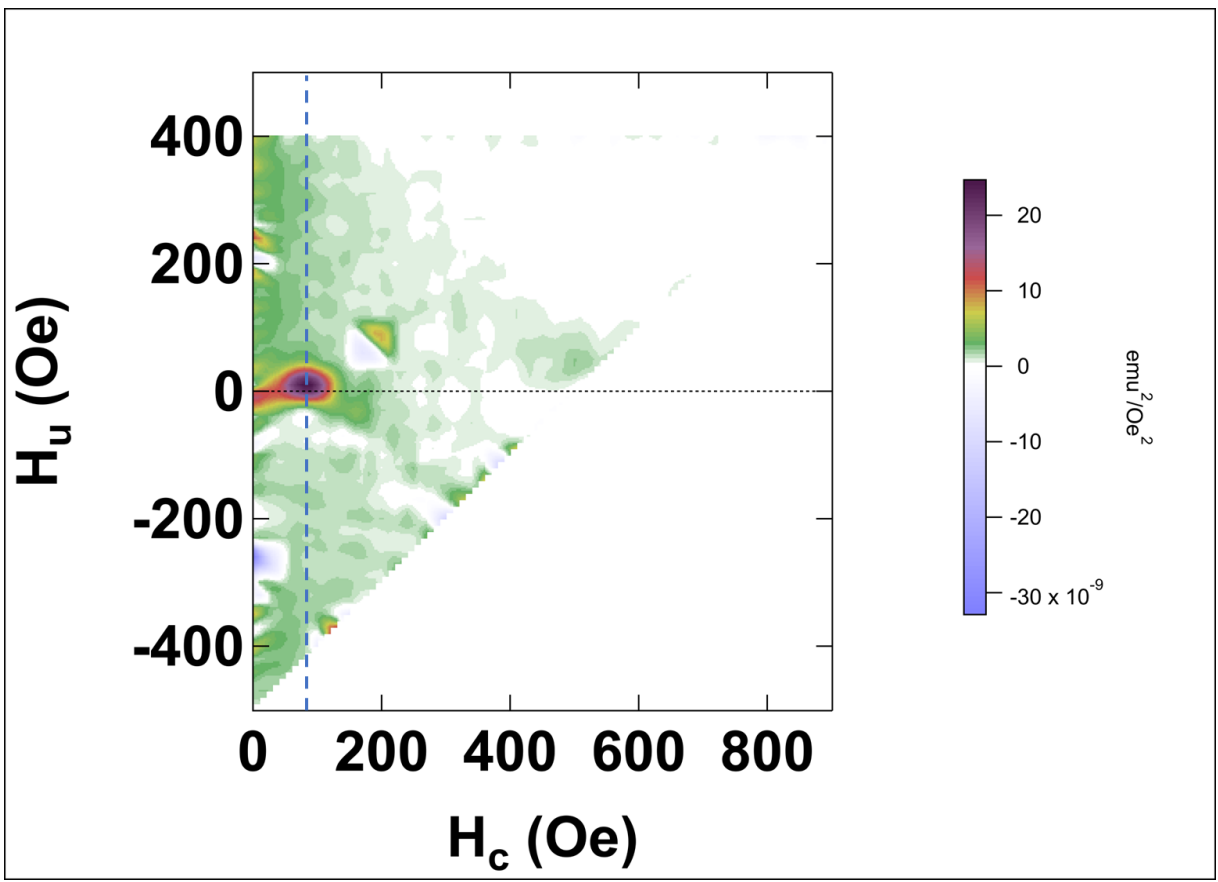

***Figure S5***. *FORC distribution for nanotubes at $\theta_H = 0°$ in $H_u$-$H_c$ coordinate.*

**References**

1. A. P. Roberts, C. R. Pike and K. L. Verosub, *Journal of Geophysical Research: Solid Earth*, 2000, **105**, 28461-28475.
2. C. R. Pike, C. A. Ross, R. T. Scalettar and G. Zimanyi, *Physical Review B*, 2005, **71**, 134407.